\documentclass[conference]{IEEEtran}
\IEEEoverridecommandlockouts
\usepackage{cite}
\usepackage{amsmath,amssymb,amsfonts}
\usepackage{algorithmic}
\usepackage{graphicx}
\usepackage{textcomp}
\usepackage{longtable}
\usepackage{xcolor}
\usepackage{url}
\usepackage{array}
\usepackage{verbatim}
\def\BibTeX{{\rm B\kern-.05em{\sc i\kern-.025em b}\kern-.08em
    T\kern-.1667em\lower.7ex\hbox{E}\kern-.125emX}}
\newcolumntype{C}[1]{>{\centering\arraybackslash}p{#1}}

\begin{document}

\newcommand{\RC}[1]{\textcolor{cyan}{\textsf{\textbf{GG}:~#1}}}
\newcommand{\TF}[1]{\textcolor{orange}{\textsf{\textbf{TF}:~#1}}}

\title{A Prototype VS Code Extension to Improve Web Accessible Development}

\author{
\IEEEauthorblockN{Elisa Calì, Tommaso Fulcini, Riccardo Coppola, Lorenzo Laudadio, Marco Torchiano}
\textit{Politecnico di Torino}\\
Turin, Italy\\
name.surname@polito.it}

\maketitle

\begin{abstract}
Achieving web accessibility is essential to building inclusive digital experiences. However, accessibility issues are often identified only after a website has been fully developed, making them difficult to address. This paper introduces a Visual Studio Code plugin that integrates calls to a Large Language Model (LLM) to assist developers in identifying and resolving accessibility issues within the IDE, reducing accessibility defects that might otherwise reach the production environment.

Our evaluation shows promising results: the plugin effectively generates functioning fixes for accessibility issues when the errors are correctly detected. However, detecting errors using a generic prompt—designed for broad applicability across various code structures—remains challenging and limited in accuracy. 

\end{abstract}

\begin{IEEEkeywords}
LLM, web accessibility, WCAG, IDE, VS Code, Accessible Web Development
\end{IEEEkeywords}

\section{Introduction}
Ensuring web accessibility is essential for inclusive access to digital content, regardless of user ability. Despite growing awareness among companies about the importance of creating accessible digital content, building and testing accessible websites and applications remains complex and time-consuming. Current automated testing tools don’t cover all the accessibility tests, leaving many to be performed manually, which often occurs when development is finished or about to be completed. B. Martins et al. \cite{accessibilityInProduction} had analyzed nearly three million web pages and they have found, on average, 30 errors per page with the 63\% of pages that had more than 10 errors. Identifying such errors as early as possible in the development phase is therefore crucial to guarantee better accessibility.

Thanks to their growing capabilities in understanding and generating code, Large Language Models (from now on, LLMs) have the potential to support practitioners throughout many different aspects of the development lifecycle. 

This paper proposes an AI-based plugin for Visual Studio Code that assists developers in identifying and resolving accessibility issues directly in the IDE (Integrated Development Environment).

\section{Background and Related Work}

\subsection{Large Language Models}
Large Language Models represent a class of deep learning models capable of understanding and generating natural language and code. Some examples are Mistral Large 2 by Mistral AI\cite{mistral}, GPT-4 by OpenAi\cite{gpt4}, Gemini by Google\cite{gemini} and Llama3 by Meta\cite{llama}.

With the growth of LLM capabilities, such as Natural Language Understanding and Natural Language Generation, they have found applications in several domains, including text generation, summarization, code generation and evaluation. 

Unlike traditional methods that often rely on rule-based approaches, LLMs offer significant advantages, such as versatility, being applicable across different tasks without modification, and efficiency, helping streamline complex processes like data analysis. Another important advantage is having a ready-to-use tool, without the need to train the model, relying on a vast amount of data.

For these reasons, there is a growing interest in applying these models to coding tasks such as checking and fixing web accessibility errors.

\begin{table*}[htbp]
    \caption{Use case features}
    \centering
    \begin{tabular}{|c|C{3cm}|C{3cm}|C{3cm}|C{3cm}|}
        \hline
        \textbf{Use case} & \textbf{Accessibility errors detected automatically}& \textbf{Errors detected by LLM}& \textbf{Report with issues and fixes}& \textbf{Fixes in the examined code file} \\
        \hline
        \textbf{FixWithAI}&  Yes& No, they are detected by the ESLint linter& Yes& Yes \\
        \hline
        \textbf{CheckAndFixWithAI}&  No, the developer has to select the code block to analyze& Yes& Yes& No, only the report is provided \\
        \hline
    \end{tabular}
    \label{tab:use-case}
\end{table*}
\subsection{Accessibility}

Accessibility refers to the \emph{ability of digital systems, services, or resources to be usable by everyone}, including people with disabilities. It ensures that all users, regardless of their physical or cognitive abilities, can interact effectively with technology. 

The Web Content Accessibility Guidelines (WCAG), developed by the W3C, serve as the global standard for web accessibility. The latest version, WCAG 2.2 \cite{wcag}, published in 2023, outlines four key principles for accessible content: 
\begin{itemize}
    \item \textbf{Perceivable}: Every user must be able to perceive the information being presented;
    \item \textbf{Operable}: Every user must be able to navigate and interact with the interface;
    \item \textbf{Understandable}: Content must be clear and easy to understand;
    \item \textbf{Robust}: Content should be compatible with current and future assistive technologies.
\end{itemize}

Testing for accessibility can be manual, automated, or involve real users with disabilities \cite{ara2024accessibility}. Manual and user testing, however, often occurs later in development, leading to higher costs for fixing issues. Integrating accessibility checks earlier in the development process, using AI-based tools or LLMs, helps reduce errors and create more accessible products from the start. This approach ensures that accessibility is addressed as an integral part of development. 

\subsection{Related work}
Several studies have explored the capabilities of LLMs in identifying code errors, including accessibility issues in web pages. One significant research is the one by López et al.\cite{lopez2024turning}. This study aimed to evaluate if LLMs could assess the correctness of three WCAG criteria that usually require manual evaluation due to the need for contextual understanding — something many automated tools lack. This study compared the results of tests that were performed using both LLMs and well-known automated analysis tools. The findings revealed that automated tools often pass tests that would fail under expert manual review. In contrast, using tailored prompt techniques, LLMs correctly identified accessibility issues that usually require manual evaluation (87\% of the applicable test cases) by evaluating the context and meaning of the application.  Despite being limited to just three WCAG principles, this research showed that LLMs could potentially automate some manual checks, offering more reliable results than current automated tools. 


I. Gur et al.\cite{gur2022understanding} examined the capabilities of LLMs in tasks such as semantic classification of HTML elements, generating descriptions for HTML inputs, and autonomous web navigation. The research highlighted the importance of model architecture, with some LLMs outperforming others, and emphasized limitations like the restricted context window\footnote{The maximum amount of input text an LLM can process at once. A limited context window can constrain the model’s understanding, as the model may lose earlier portions of long inputs, potentially reducing response quality.}, which impacted the model's understanding and performance. 

These studies can serve as a foundation for further exploration into the role of LLMs in improving accessibility testing and automating tasks that currently require manual intervention. To the best of our knowledge, this paper present the first attempt aimed at approaching an LLM directly inside an IDE, allowing  early accessibility assessment intervention to reduce post-development remediation efforts.

\section{Proposed Approach}
This paper will discuss two use cases aimed at helping developers identify and resolve accessibility issues in their code: \textbf{FixWithAI} and \textbf{CheckAndFixWithAI}. 

In {FixWithAI}, accessibility errors are {automatically detected} by the linter ESLint and developers can receive suggestions for fixes on demand. The system provides a detailed description of the issue along with possible solutions. The corrected code is then made available, in the examined code file and in a txt file, for the developer to review and implement, ensuring the accessibility issue is addressed efficiently.

{CheckAndFixWithAI}, instead, allows developers to manually select sections of their code for analysis. After choosing a specific block, the system examines it for accessibility problems, generating a report with both the identified issues and recommended fixes. This gives developers the flexibility to focus on specific parts of their code, receiving targeted feedback and solutions.

In Table \ref{tab:use-case} all features divided by use cases are reported.

Both use cases streamline the process of diagnosing and correcting accessibility problems directly {evaluating the code inside the IDE}, enabling developers to enhance their code with minimal effort and clear guidance.
These two use cases were designed to be lightweight and be available on demand, avoiding to oveload the model with continuous requests.

\section{Methodology}
This chapter outlines the methodology used to implement the two use cases, {FixWithAI} and {CheckAndFixWithAI}, focusing on the LLMs evaluated, the use of ESLint, the prompt engineering process, and the overall architecture that supports the system.

\subsection{ESLint linter}
ESLint \cite{eslint} is a popular linter, i.e. a source code error/vulnerabilities detector, for JavaScript. It is commonly used to statically identify code issues.
It supports several plugins, including {eslint-plugin-jsx-a11y} \cite{jsxally}, a plugin that checks accessibility.
This plugin, which can successfully identify accessibility issues with static analysis, all alone has limited ability to fix issues, with trivial and non-context-dependent suggestions.

Therefore, we decided to use eslint-plugin-jsx-a11y as a basis for identifying errors in our FixWithAI mode.
We configured ESLint, with its accessibility plugin to detect the shortcomings and added in the \emph{Quick Fix} menu the \emph{fixAccessibilityIssue} entry.
By relying on that, the developer decides to ask the AI for suggestions on how to solve the issue. Although, ES Linter has limited ability to identify complex accessibility issues, being more effective in identifying simpler problems. Therefore, we have designed the CheckAndFix usage mode precisely to make up for the identification deficits of static analysis.




\subsection{LLMs utilized}
For this project, four LLMs were evaluated: \emph{Llama2, Llama3, CodeLlama, CodeGemma}\cite{ollama}. They were chosen due to their advanced natural language understanding, because they are free to use and, for the last two, for their code interpretation capabilities. To choose the LLM to be adopted, we performed a qualitative evaluation by feeding a set of ten known accessibility problems with the same prompts and assessing the responses of the different models. We chose the model based on the number of accessibility problems correctly corrected by the LLM.
As a result of the evaluation, we chose \emph{CodeLlama} for the implementation of the two use cases.

\subsection{PromptMixer and Prompt Engineering}
A key part of this methodology involved optimizing the prompts used to query the LLM. To reach this goal, the tool \emph{PromptMixer} \cite{promptmixer} was used. It helped to build and compare versioned prompts to maximize the LLM's output quality. We used it also to feed the LLMs with the fixed set of prompts in the selection step to facilitate the selection of the most suitable model. It allowed us to simultaneously compare LLMs responses, including Llama2, Llama3, and CodeLlama. However, being CodeGemma unsupported by PromptMixer at the time of writing, we had to manually validate it only using its API.

In the FixWithAI use case, accessibility errors are automatically detected by ESLint. After an error has been detected, the model intervenes by receiving a prompt that includes the faulty code snippet and a list of error descriptions identified by ESLint. By clicking the contextual menu the user can ask for help from the AI assistant: a prompt requests the model to fix the code in question. The model will then generate a list of possible solutions, which will be displayed below the flawed code. Additionally, these solutions will be provided in a separate text file for further reference.

In the {CheckAndFixWithAI} use case, the system has to first identify accessibility errors and then fix them. For this reason, we relied on prompt chaining~\cite{wu2022promptchainer}: the first prompt aims to find accessibility errors and return it as a list of objects reporting a description of the error, the part of the code where the error is present, and the WCAG criteria number that is violated; the second prompt takes as input the response of the first LLM interrogation and return the original list of objects integrated with a description of the resolution and the original code fixed. When selecting the desired piece of code, a new entry appears in the context menu, where the user can activate the use case.

In both cases, we adopted the role prompting technique, giving the model the role of a front-end developer with web accessibility knowledge~\cite{zhang2023visar}. It should assess JavaScript and React.js applications to verify that all the code is WCAG 2.2 compliant. Moreover, to avoid inaccuracy, the prompts specified the expected result template. An example of the prompt used can be found in the online appendix\footnote{\url{https://doi.org/10.6084/m9.figshare.27636123.v1}}

\subsection{System architecture}
The architecture consists of two primary components: ESLint for static accessibility error detection and LLM for generating analysis and fixing suggestions. ESLint must be configured within the front-end project to perform its role effectively. The extension supports JavaScript, TypeScript, and React-based code (JSX and TypeScript). 

For the use case FixWithAI, once ESLint identifies an issue, whether the developer decides to query the LLM, it generates a possible correction based on an optimized prompt. The result is shown under the code analyzed. 

For the other use case, CheckAndFixWithAI, the code chosen by the developer is sent to the LLM. This will analyze the code to find accessibility issues and, if needed, fix them.

Both use cases generate file results that can be saved by the developer in a persistent way for future reference. This architecture allows developers to address accessibility issues dynamically, combining real-time static analysis with the LLM’s adaptive capabilities.

\begin{figure}
    \centering
    \includegraphics[width=1\linewidth]{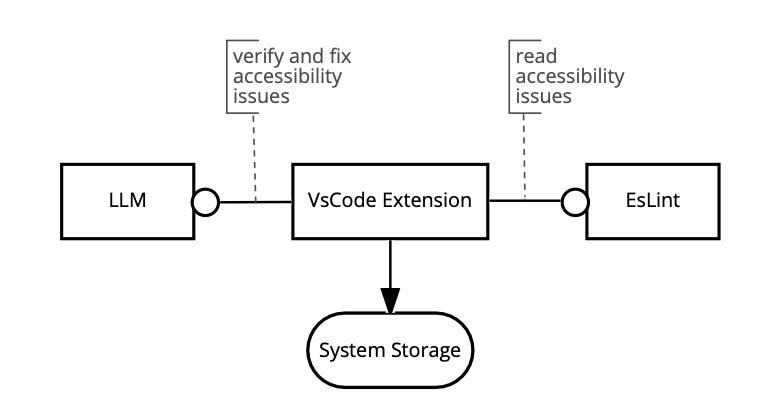}
    \caption{System architecture diagram}
    \label{fig:sistem-architecture-diagram}
\end{figure}

Figure \ref{fig:sistem-architecture-diagram} describes the architecture of the system and all the components involved.

\section{The prototype}
In this chapter, two examples are presented to illustrate how the system works in practice, showing how accessibility issues in the code are detected and resolved using the integrated tools. The code that was evaluated is part of a React.js component.

\subsection{Use Cases}
\subsubsection{Use Case 1: FixWithAI}
In this scenario, the developer opens a file containing React.js code with potential accessibility issues that don't need context to be detected. ESLint immediately detects these issues through static analysis. Let’s pretend that the developer opens the file Tooltip.js. Inside it, there is a \textit{div} element with an \textit{onClick} event. This element is not accessible to people navigating websites via assistive technologies. To make it accessible, it should be implemented as \textit{button} or it should have the “button” role and all the keyboard interaction should be handled. Due to these lacks, ESLint flags the underlined as an error. In this case, the developer is not sure of the right solution, so they hover over the flagged issue and, through the quick fix menu, select the "Get fix suggestion from AI" option to get a fix suggestion. 

\begin{figure}
    \centering
    \includegraphics[width=\linewidth]{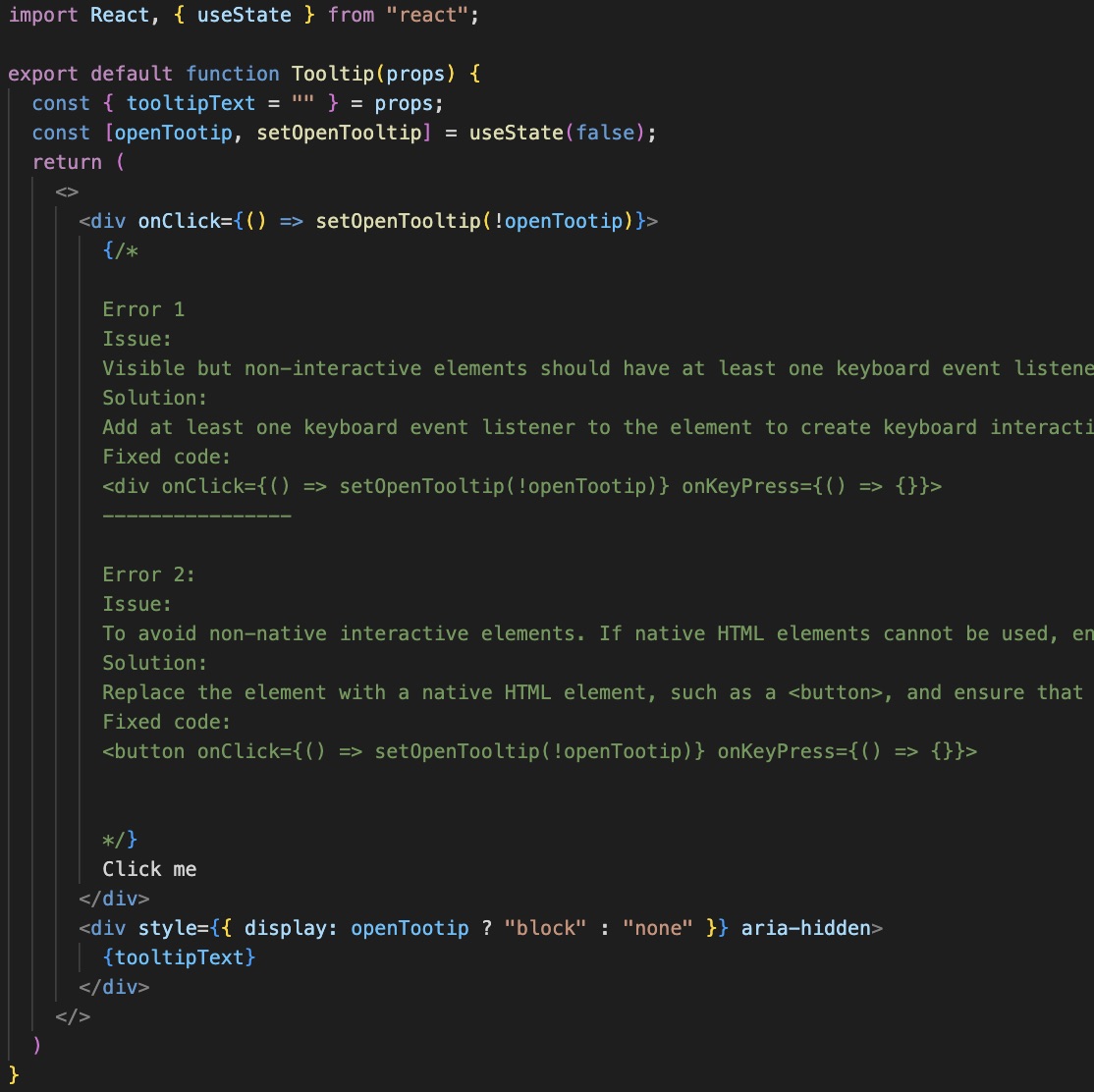}
    \caption{Fix suggested by the LLM inside the code file}
    \label{fig:llm-report-file}
\end{figure}

At this point, the LLM analyzes the flagged issue and returns a possible fix. 
Figure \ref{fig:llm-report-file} shows how the response of the LLM is reported under the code that generated the accessibility issue. In this way, the developer can see the correction in the context of the file. 

\subsubsection{Use Case 2: CheckAndFixWithAI}
In the second use case, the developer wants to check the accessibility of a block of code. A possible scenario is when the developer wants to check the accessibility of a block of code in the context of a bigger block of code. So, they select a block of code and, invoking the "Check and fix accessibility issues with AI" command, the developer submits the selected code to the LLM for analysis.

At the end of the query, a textual file is created. It is composed by three sections:
\begin{itemize}
    \item \textit{Code section} which reports the code analyzed by the LLM. It corresponds to the developer selection.
    \item \textit{Errors section} which lists the accessibility errors detected by the LLM, the source code generating them and the violated WCAG criteria.
    \item \textit{Fixes sections} contains the same information as the error section, with the addition of the description of the fix strategy and the fixing code.
\end{itemize}

\begin{figure}
    \centering
    \includegraphics[width=0.75\linewidth]{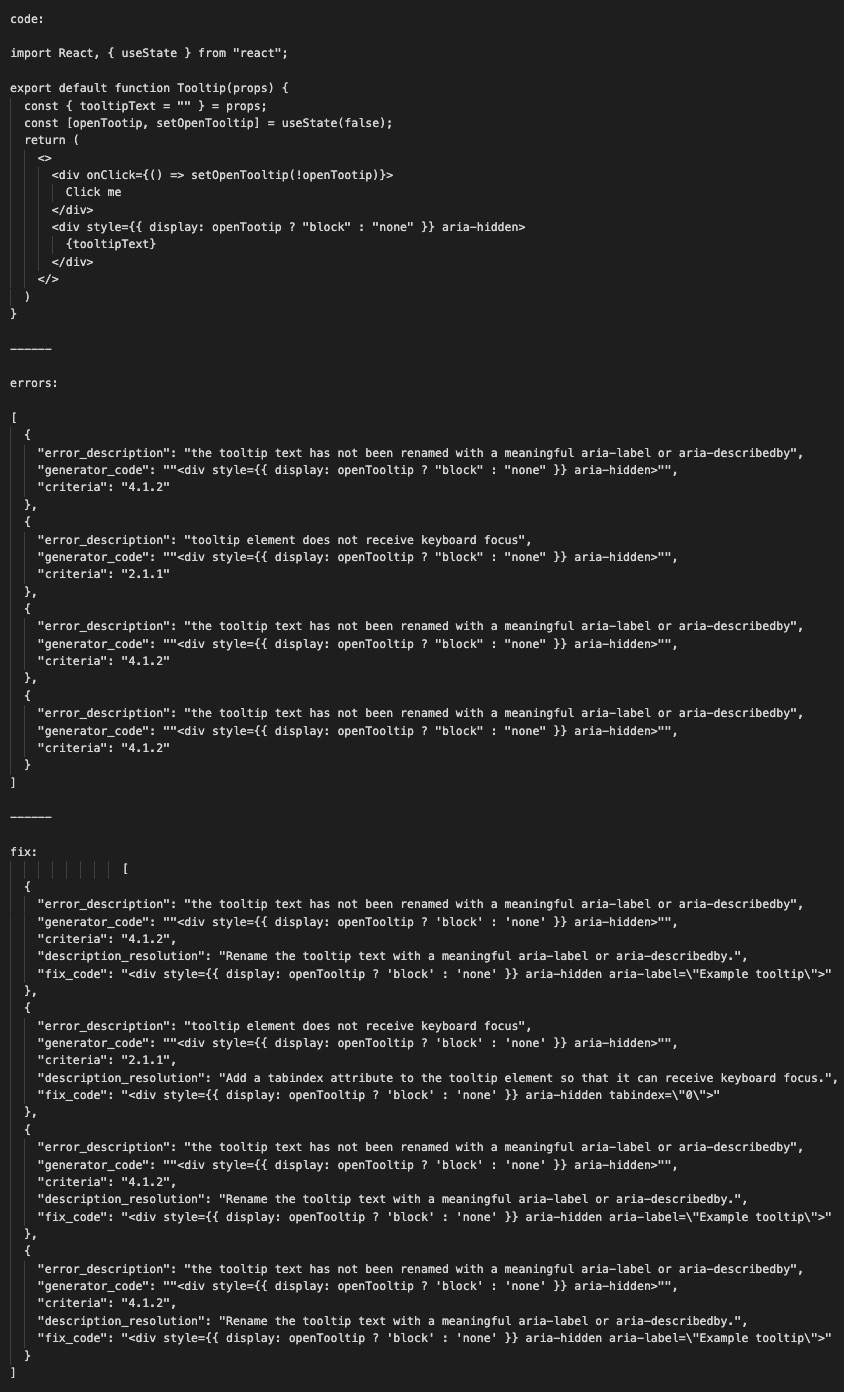}
    \caption{Response of the CheckAndFixWithAI use case}
    \label{fig:response-check-fix}
\end{figure}

If we interrogate the LLM with the same code of the \textbf{FixWithAI} use case, four errors are detected (Figure \ref{fig:response-check-fix}).

\subsection{Evaluation}
The results obtained during the testing phase, in terms of responses provided by the LLM, are classified into three categories (\textit{Correct}, \textit{Partially correct}, and \textit{Incorrect})
The evaluation process consists of a manual assessment of the responses received from the LLM according to the defined criteria.
The classification criteria follow a structured process: first, we check if any “Incorrect” criteria are satisfied. If none of them are met, the result is checked for "Partially correct". When the majority of the "Partially correct" criteria are met, the result is considered "Correct". This is because, even if it is not the most exhaustive, the solution is still adequate as it does not alter the functionality while partially resolving the accessibility error.

Each use case requires specific criteria. The criteria for the evaluation with the categories for each use case are reported in a table in the online appendix\footnote{\url{https://doi.org/10.6084/m9.figshare.28194347}}.
At least three of the authors participated in the evaluation of each response received by the LLM. Each one used the criteria table as a checklist, then the different responses were compared and analyzed. In case of disagreements, a discussion was planned, with a final decision based on majority voting. After all the discussions, all the authors agreed in all the classifications.

\subsubsection{FixWithAI evaluation}
In the FixWithAI use case, the LLM had to fix two errors identified by ESLint: an interactive element must have keyboard handling. It is preferable to avoid placing interaction listeners in non-interactive elements. Following the classification criteria described, the LLM response (Figure \ref{fig:llm-report-file}) results as correct as it satisfies 5 criteria of the "Partially correct" level. Indeed, even though the code presents more than two errors which are not detected by ESLint (such as the absence of the "tooltip" role, the aria-label/aria-labelledby/aria-describedby that refer to the tooltip description and the aria-hidden on the tooltip text).

We acknowledge that the solution to error 2 reports the onKeyPress function on a button element, considered useless since the button already handles all the keyboard interactions. Moreover, the solution to error 1 is incomplete because also the onKeyUp event listener is needed to handle interaction with the spacebar\cite{button_wcag}.

\subsubsection{CheckAndFixWithAI evaluation}
In the CheckAndFixWithAI use case (Figure \ref{fig:response-check-fix}) the LLM identified 4 errors. Unfortunately, 3 of them are repeated, while other accessibility errors listed in the previous paragraph were not identified and an aria-label is exposed on an aria-hidden element. For these reasons, this proposed solution is classified as "Incorrect". Therefore, we believe that the prototype requires refinement of the prompts used for error detection and correction to better analyze large chunks of code.

\subsection{Threats to Validity}

This paper present a prototype extension for VS code. The prototypical nature of the tool is, inherently affected by limitations, which we list here and will be addressed in future development iteration. A first threat to validity is the programming language used as object. React, indeed, is one of the most widespread framework for web development~\cite{react}, although to reach a comprehensive audience, the extension should receive further support for other framework such as Express, Angular or Vue.

Another possible weakness may be the prompt engineering techniques used and the LLMs selected. We envision to study in a future study how different prompt engineering techniques influences the accesibility issue detection and correction capability of the different LLMs. Effectiveness will not be the only aspect to be addressed in such a comparative study, but we will also consider other non-functional aspects such as the time it takes the model to respond to a prompt and the amount of tokens required for the interactions, thus the monetary cost whether an LLM available under subscription is chosen.

Finally, the evaluation provided is totally qualitative, this means that these results can be considered partly subjective. To mitigate this problem, we relied on different points of view and discussions. However, the objective of the presented evaluation is not intended to be comprehensive, rather than just a way to estimate the goodness of the developed prototype. Once the tool reaches a greater degree of maturity, it is intended to conduct a full-fledged empirical validation to corroborate the research with trustable results.

\section{Conclusion and Future Work}

This paper introduced a VS Code extension that uses LLMs to help developers detect and fix web accessibility issues within the development environment, enhancing the accessibility workflow through two primary use cases: \textit{FixWithAI} and \textit{CheckAndFixWithAI}. The FixWithAI use case allows developers to receive AI-generated accessibility fixes for errors detected by ESLint. In contrast, the CheckAndFixWithAI allow the developer to detect not signaled accessibility issues in specific code sections. 

Through these two approaches, the plugin helps in finding accessibility issues earlier in the development process, minimizing manual efforts and improving code quality.
State-of-the-art tools either provide static detection with no context-dependent suggestions or an evaluation of the finished product, not in the development phase. Conversely, we provide an approach to dynamically detect and solve accessibility issues inside the coding environment.

\subsection{Future work}
A key next step is to conduct a comprehensive evaluation of the plugin's functionality, accuracy and usability. 

Preliminary results for the FixWithAI use case are promising, receiving good fix suggestions. Conversely, the CheckAndFixWithAI use case requires further refinement as some issues, such as a repeated listing of the same error (Figure \ref{fig:response-check-fix}), badly affect its effectiveness. 
Addressing this shortcoming will involve optimizing the prompt design to reduce redundancy in error identification and improve clarity in responses. We envision this adjustment will enhance the overall precision and usability of the CheckAndFixWithAI functionality. CheckAndFixWithAI could benefit from advanced prompt engineering techniques to improve error detection accuracy and eliminate redundancies in error reporting. To this extent we plan to run a comparative study of the capability of that use cases, when adopting different prompt engineering techniques.

Further refinement could include enhancing the plugin's adaptability and configurability, allowing customization to meet specific project requirements. 
This may involve options for tailored output formats or more granular control over accessibility compliance criteria.

Additionally, as future development efforts of the plugin, it could be extended to support more programming languages and frameworks such as Vue.js, Angular, and Svelte as currently, the plugin focuses on JavaScript, TypeScript, and React-based code. This addition will allow for wider adoption across diverse projects.

\bibliographystyle{IEEEtran}
\bibliography{bibliography.bib}

\end{document}